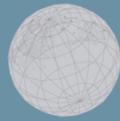

# 23rd International Conference on Science and Technology Indicators
## "Science, Technology and Innovation Indicators in Transition"

**STI 2018 Conference Proceedings**

*Proceedings of the 23rd International Conference on Science and Technology Indicators*

All papers published in this conference proceedings have been peer reviewed through a peer review process administered by the proceedings Editors. Reviews were conducted by expert referees to the professional and scientific standards expected of a conference proceedings.

**Chair of the Conference**

Paul Wouters

**Scientific Editors**

Rodrigo Costas
Thomas Franssen
Alfredo Yegros-Yegros

**Layout**

Andrea Reyes Elizondo
Suze van der Luijt-Jansen

The articles of this collection can be accessed at https://hdl.handle.net/1887/64521

ISBN: 978-90-9031204-0

© of the text: the authors

© 2018 Centre for Science and Technology Studies (CWTS), Leiden University, The Netherlands

# Testing Reviewer Suggestions Derived from Bibliometric Specialty Approximations in Real Research Evaluations

Nadine Rons[*]

[*]*Nadine.Rons@vub.ac.be*
Research and Data Management Department, Vrije Universiteit Brussel, Pleinlaan 2, BE-1050 Brussels (Belgium)

**Abstract**
Many contemporary research funding instruments and research policies aim for excellence at the level of individual scientists, teams or research programmes. Good bibliometric approximations of related specialties could be useful for instance to help assign reviewers to applications. This paper reports findings on the usability of reviewer suggestions derived from a recently developed specialty approximation method combining key sources, title words, authors and references (Rons, 2018). Reviewer suggestions for applications for Senior Research Fellowships were made available to the evaluation coordinators. Those who were invited to review an application showed a normal acceptance rate, and responses from experts and coordinators contained no indications of mismatched scientific focus. The results confirm earlier indications that this specialty approximation method can successfully support tasks in research management.

**Data and Method**
For six applicants for a 5-year Senior Research Fellowship term awarded by the Vrije Universiteit Brussel, specialty approximations were constructed as described by Rons (2018). This method's approach is founded in concepts defining research disciplines (Sugimoto and Weingart, 2015) and scholarly communication (Ni, Sugimoto, and Cronin, 2013), and in empirically observed regularities for sources (Bradford, 1934; Garfield, 1971), title words (Zipf, 1935), authors (Lotka, 1926) and references (Price, 1965):
- Phase 1: Construction of the seed record, for individual scientists by enlarging the publication record with the publications referred to.
- Phase 2: Determination of the most frequently occurring values ('key values') among cells of sources (Rons, 2014), title words, authors and references, required to 'cover' at least a predefined majority ('coverage threshold') of publications in the seed record.
- Phase 3: Identification of all publications covered by key values for at least three of the four data fields, constituting the specialty approximation.

In all phases the same operationalizations were used as by Rons (2018), except for a lower coverage threshold of 50% instead of 80% in phase 2. This setting more strongly focuses results on the most representative field values, which is more important than exhaustivity in this application where the only aim is to select prominent scientists in the specialty as potential reviewers. In phases 1 and 3, publications were collected from the online Web of Science of Clarivate Analytics (key figures in Table 1).

Table 1. Key figures for the applicants' publication records, seed records and specialty approximations.

| $A_i$ | $A_1$ | $A_2$ | $A_3$ | $A_4$ | $A_5$ | $A_6$ |
|---|---|---|---|---|---|---|
| $D_i$ | Bioengineering | Bioengineering | Bioengineering | Biology | Engineering | Engineering |
| $PR_i$ | 11 | 26 | 26 | 10 | 86 | 29 |
| $SR_i$ | 44 | 82 | 126 | 29 | 164 | 87 |
| $SA_i$ | 147 | 100 | 2291 | 39 | 189 | 403 |

$A_i$: Applicant $i$.
$D_i$: Domain in which the specialty of $A_i$ is situated.
$PR_i$, $SR_i$ and $SA_i$: Numbers of publications in respectively the publication record, seed record and specialty approximation for $A_i$.
Publication period: 2014-2017, until dates specified below.
*Data sourced from Clarivate Analytics (formerly Thomson Reuters' IP & Science business). Web of Science accessed online 11 September 2017 (publication records), 28 September 2017 (seed records), 3 October - 4 December 2017 (specialty approximations).*

From each specialty approximation five to seven prominent authors were suggested as potential reviewers, starting from those with the highest numbers of publications, and excluding co-authors of the applicant, authors with a same institutional affiliation, authors with a low professional grade compared to that of the applicant, reviewer suggestions submitted by the applicant, and authors with whom a collaboration is mentioned in the application. Aiming to obtain at least three reviews per application, the academic evaluation coordinators selected about twice as many potential reviewers, partly from the suggestions submitted by the applicant and partly not from these suggestions. For the latter, they were free to (partly) use or not to use the 'bibliometric suggestions' derived from the specialty approximations.

**Results & Discussion**

Indications of success of the bibliometric method in suggesting reviewers are situated in two phases (overview of results in Table 2):

1. Usage by the coordinator: For 5 of the 6 applications, the coordinators chose to (partly) use the bibliometric suggestions. In total, of the 18 experts invited to review who were not from the applicants' suggestions, 8 were bibliometric suggestions.
2. Response to invitations: Of the 8 bibliometric suggestions invited to review, 4 accepted the invitation and 4 declined it or did not respond. This is a similar acceptation rate as for the invitations sent to all other experts (28), of whom also about half (15) accepted the invitation and about half (13) declined it or did not respond.

Reasons given by experts for a declined invitation or by evaluation coordinators for non-usage of certain bibliometric suggestions contained no indications of a mismatched scientific focus.

Table 2. Selection of potential reviewers and response to invitations.

| 1. Selection of potential reviewers | | 2. Response to invitations to review | | |
|---|---|---|---|---|
| Total | Invited to review | Accepted | Declined | No response |
| [a] Suggestions by the applicants (*) | | | | |
| 30(12) | 18(5) | 10(2) | 3(0) | 5(3) |
| [b] Bibliometric suggestions derived from the applicants' specialty approximations | | | | |
| 34 | 8 | 4 | 1 | 3 |
| [c] Experts identified by the evaluation coordinators, not from [a] nor from [b] (*) | | | | |
| 11(3) | 10(2) | 5(1) | 3(0) | 2(1) |
| (*) Between brackets: numbers present as authors in the applicants' specialty approximations | | | | |

**Discussion and Conclusions**

While several systems exist to assign submissions to reviewers from a given set, or to find reviewers using text mining, to the best of the author's knowledge this is the first report in scientific literature on the usability and performance of ad hoc bibliometrically generated reviewer suggestions in a real evaluation context.

The cases included in this paper are limited in number, but nevertheless broaden the range of reported results of the recently introduced specialty approximation method towards specialties in the domains of Engineering and Bioengineering. It is shown that, as a source for reviewer suggestions, these new specialty approximations can be as adequate as customary approaches for the type of evaluation concerned. These results confirm the method's potential for usage in evaluation contexts — where operationalization options can be further explored — and its potential interest for other types of applications.

**References**

Bradford, S.C. (1934). Sources of Information on Specific Subjects. *Engineering: An Illustrated Weekly Journal*, 137(3550), 85-86.

Garfield, E. (1971). The Mystery of the Transposed Journal Lists -- Wherein Bradford's law of Scattering is Generalized According to Garfield's Law of Concentration. *Current Contents*, 17, 5-6.

Lotka, A.J. (1926). The frequency distribution of scientific productivity. *Journal of the Washington Academy of Sciences*, 16(12), 317-323.

Ni, C., Sugimoto, C.R., & Cronin, B. (2013). Visualizing and comparing four facets of scholarly communication: Producers, artifacts, concepts, and gatekeepers. *Scientometrics*, 94(3), 1161-1173

Price, D.J. (1965). Networks of Scientific Papers. The pattern of bibliographic references indicates the nature of the scientific research front. *Science*, 149(3683), 510-515.

Rons, N. (2018). Bibliometric approximation of a scientific specialty by combining key sources, title words, authors and references. *Journal of Informetrics*, 12(1), 113–132.

Rons, N. (2014). Investigation of Partition Cells as a Structural Basis Suitable for Assessments of Individual Scientists. In: *Proceedings of the science and technology indicators conference 2014 Leiden "Context Counts: Pathways to Master Big and Little Data"*, 3-5 September 2014, Leiden, the Netherlands, Ed Noyons (Ed.), 463-472.

Sugimoto, C.R., & Weingart, S. (2015). The kaleidoscope of disciplinarity. *Journal of Documentation*, 71(4), 775-794.

Zipf, G.K. (1935). *The psycho-biology of language: An introduction to dynamic philology* (Boston: Houghton Mifflin).